\documentclass[pdflatex,sn-mathphys]{sn-jnl}

\jyear{2021}%

\theoremstyle{thmstyleone}%
\theoremstyle{thmstyletwo}%

\theoremstyle{thmstylethree}%

\usepackage{subcaption}
\usepackage[section]{placeins}
\usepackage{tabularx}
\usepackage{multirow}
\usepackage{CJKutf8}

\usepackage{float}
\usepackage{etoolbox}
\makeatletter
\patchcmd{\ps@headings}
{\hbox to \hsize{\hfill Springer Nature 2021 \LaTeX\ template\hfill}}
{\hbox to \hsize{}}
{}
{}
\patchcmd{\ps@titlepage}
{\hbox to \hsize{\hfill Springer Nature 2021 \LaTeX\ template\hfill}}
{\hbox to \hsize{}}
{}
{}
\makeatother
\makeatletter
\patchcmd{\ps@headings}
{\hbox to \hsize{\hfill Springer Nature 2021 \LaTeX\ template\hfill}}
{\hbox to \hsize{}}
{}
{}
\patchcmd{\ps@headings}
{\hbox to \hsize{\hfill Springer Nature 2021 \LaTeX\ template\hfill}}
{\hbox to \hsize{}}
{}
{}
\patchcmd{\ps@titlepage}
{\hbox to \hsize{\hfill Springer Nature 2021 \LaTeX\ template\hfill}}
{\hbox to \hsize{}}
{}
{}
\makeatother

\begin{document}

\begin{CJK*}{UTF8}{gbsn}

\title[FuXi: 15-day global weather forecasts]{FuXi: A cascade machine learning forecasting system for 15-day global weather forecast}


\author[1]{\fnm{Lei} \sur{Chen}}\email{cltpys@163.com}
\equalcont{These authors contributed equally to this work.}
\author[1]{\fnm{Xiaohui} \sur{Zhong}}\email{x7zhong@gmail.com}
\equalcont{These authors contributed equally to this work.}

\author[2,3]{\fnm{Feng} \sur{Zhang}}\email{fengzhang@fudan.edu.cn}

\author[1]{\fnm{Yuan} \sur{Cheng}}\email{cheng$\_$yuan@fudan.edu.cn}

\author[1]{\fnm{Yinghui} \sur{Xu}}\email{renji.xyh@vip.163.com}

\author*[1]{\fnm{Yuan} \sur{Qi}}\email{qiyuan@fudan.edu.cn}

\author*[1]{\fnm{Hao} \sur{Li}}\email{lihao$\_$lh@fudan.edu.cn}

\affil[1]{\orgdiv{Artificial Intelligence Innovation and Incubation Institute}, \orgname{Fudan University}, \orgaddress{\city{Shanghai}, \postcode{200433}, \country{China}}}

\affil[2]{\orgdiv{Key Laboratory of Polar Atmosphere-ocean-ice System for Weather and Climate, Ministry of Education, Department of Atmospheric and Oceanic Sciences}, \orgname{Fudan University}, \orgaddress{\city{Shanghai}, \postcode{200433}, \country{China}}}

\affil[3]{\orgname{Shanghai Qi Zhi Institute}, \orgaddress{\city{Shanghai}, \postcode{200232}, \country{China}}}

\abstract{Over the past few years, the rapid development of machine learning (ML) models for weather forecasting has led to state-of-the-art ML models that have superior performance compared to the European Centre for Medium-Range Weather Forecasts (ECMWF)'s high-resolution forecast (HRES), which is widely considered as the world's best physics-based weather forecasting system. Specifically, ML models have outperformed HRES in 10-day forecasts with a spatial resolution of 0.25\textdegree. However, the challenge remains in mitigating accumulation of forecast errors for longer effective forecasts, such as achieving comparable performance to the ECMWF ensemble in 15-day forecasts. Despite various efforts to reduce accumulation errors, such as implementing autoregressive multi-time step loss, relying on a single model has been found to be insufficient for achieving optimal performance in both short and long lead times. Therefore, we present FuXi, a cascaded ML weather forecasting system that provides 15-day global forecasts at a temporal resolution of 6 hours and a spatial resolution of 0.25\textdegree. FuXi is developed using 39 years of the ECMWF ERA5 reanalysis dataset. The performance evaluation demonstrates that FuXi has forecast performance comparable to ECMWF ensemble mean (EM) in 15-day forecasts. FuXi surpasses the skillful forecast lead time achieved by ECMWF HRES by extending the lead time for ${Z500}$ from 9.25 to 10.5 days and for ${T2M}$ from 10 to 14.5 days. Moreover, the FuXi ensemble is created by perturbing initial conditions and model parameters, enabling it to provide forecast uncertainty and demonstrating promising results when compared to the ECMWF ensemble.}


\keywords{weather forecast, machine learning, accumulation error, cascade, FuXi, transformer}

\maketitle

\section{Introduction}\label{sec1}

Accurate weather forecasts play an important role in many aspects of human society. Currently, national weather centers around the world generate weather forecasts using numerical weather prediction (NWP) models, which simulate the future state of the atmosphere. Nevertheless, running NWP models often requires high-performance computing systems, with some simulations taking several hours using thousands of nodes. The Integrated Forecast Systems (IFS) of the European Centre for Medium-range Weather Forecast (ECMWF) is widely regarded as the most accurate global weather forecast model \citep{ECMWF2021}. The ECMWF's high-resolution forecast (HRES) runs at a horizontal resolution of 0.1\textdegree with 137 vertical levels for 10-day forecasts. However, uncertainty in weather forecasts is inevitable due to the limited resolution, approximation of physical processes in parameterizations, errors in initial conditions (and boundary conditions for regional models), and the chaotic nature of the atmosphere. Additionally, the degree of uncertainty and the magnitude of errors in weather forecasts increases as forecast lead time. One way to address this uncertainty is to run an ensemble of forecasts by incorporating perturbations in initial conditions and physical parameterizations in the NWP model. The ECMWF ensemble prediction system (EPS) \cite{Magnusson2019} provides forecasts up to 15 days and is comprised of one control member and 50 perturbed members. The IFS Cycle 48r1, which was introduced in June 2023, upgrade the spatial and vertical resolution and vertical resolution of the EPS same as the HRES \cite{balsamo2023recent}, which was made possible by a new supercomputer with enhanced capacity. Prior to the upgrade, the EPS ran at a lower spatial resolution of 18 km and had fewer vertical levels of 91, due to the substantial computational demands of running 51 members with limited computing resources. 


In recent years, there have been increasing efforts to replace the traditional NWP models with machine learning (ML) models for weather forecasting \cite{Schultz2021}. ML-based weather forecasting systems have several advantages over NWP models, including faster speeds and the potential to provide higher accuracy than uncalibrated NWP models due to training with reanalysis data \cite{Schultz2021}. To facilitate intercomparison between different ML models, the WeatherBench benchmark was introduced to evaluate medium-range weather forecasting (i.e., 3-5 days) \cite{rasp2020weatherbench,garg2022weatherbench}. WeatherBench was created by regridding ERA5 reanalysis data \cite{hersbach2020era5} from $0.25^{\circ}$ resolution to three different resolutions ($5.625^{\circ}$, $2.8125^{\circ}$ and $1.40625^{\circ}$). Several studies have aimed to improve forecast performance on this dataset \cite{rasp2021data, weyn2020improving, hu2022swinvrnn}. For example, Rasp et al. \cite{rasp2021data} used a deep residual convolutional neural network (CNN) known as ResNet \cite{he2016deep} to predict 500 hPa geopotential (${Z500}$), 850 hPa temperature (${T850}$), 2-meter temperature (${T2M}$), and total precipitation (${TP}$) at a spatial resolution of $5.625^{\circ}$ for up to 5 days. They found that the ResNet model has similar performance compared to physical baseline models, such as IFS T42 and T63, with a comparable resolution. Meanwhile, Hu et al. \cite{hu2022swinvrnn} proposed the SwinVRNN model, which utilizes a Swin Transformer-based recurrent neural network (RNN) (SwinRNN) model coupled with a perturbation module to learn multivariate Gaussian distributions based on the Variational Auto-Encoder framework. They demonstrated the SwinVRNN model's potential as a powerful ML-based ensemble weather forecasting system, with good ensemble spread and better accuracy compared to IFS in terms of ${T2M}$ and 6-hourly ${TP}$ in 5-day forecasts with a $5.625^{\circ}$ resolution. 

While ML models have shown good performance in weather forecasting, their practical values are limited because of their forecasts' low resolution (e.g., $5.625^{\circ}$). As a remarkable breakthrough, the FourCastNet model \cite{pathak2022fourcastnet} is the first of its kind to provide high-resolution global weather forecasts of $0.25^{\circ}$ for a time period of 7 days. It integrates the Adaptive Fourier neural operator (AFNO) \cite{guibas2022adaptive} with a Vision Transformer (ViT) \cite{ViT2021}. However, FourCastNet's forecast accuracy is still worse than HRES's. SwinRDM \cite{swinRDM2023} distinguishes itself as the first ML-based weather forecasting system to outperform ECMWF HRES in 5-day forecasts at a spatial resolution of $0.25^{\circ}$. SwinRDM integrates SwinRNN+, an improved version of SwinRNN that surpasses ECMWF HRES at a spatial resolution of $1.40625^{\circ}$, with a diffusion-based super-resolution model that increases the resolution to $0.25^{\circ}$. Pangu-Weather \cite{bi2022panguweather} shows its superior performance compared to ECMWF HRES in 7-days forecasts at a resolution of $0.25^{\circ}$. Additionally, GraphCast \cite{lam2022graphcast}, an autoregressive model that implements a graph neural network (GNN), outperforms HRES in 90${\%}$ of the 2760 variable and lead time combinations in 10-day forecasts.


Although ML models have shown promising results in generating weather forecasts for 10 days, long-term forecasting remains challenging due to cumulative errors. The iterative forecasting method, which uses the model outputs as inputs for subsequent predictions, is a commonly used approach in developing ML-based weather forecasting systems. This approach is similar to the time-stepping methods used in conventional NWP models \cite{Dueben2018}. However, as the number of iterations increases, errors in the model outputs accumulate, which may lead to significant discrepancies with the training data and unrealistic values in long-term forecasts. Many research has been conducted to enhance the stability and accuracy of long-term forecasts. Weyn et al. \cite{weyn2020improving} proposed a multi-time-step loss function to minimize errors over multiple iterated time steps. Rasp et al. \cite{rasp2020weatherbench} compared iterative forecasts with direct forecasts that predict specific lead times and found the latter to be more accurate. However, one limitation of direct forecasts is that separate models need to be trained for each lead time. The FourCastNet \cite{pathak2022fourcastnet} model underwent two training phases: pre-training, in which the model is optimized to map one time step to the next with a 6-hour interval, and fine-tuning to minimize errors in two-step prediction, similar to the multi-time-step loss function proposed by Weyn et al. On the other hand, Bi et al. proposed a hierarchical temporal aggregation strategy for Pangu-Weather's forecasts, training four separate models for 1-hour, 3-hour, 6-hour, and 24-hour forecasts \cite{bi2022panguweather}. They demonstrated that running the 24-hour model 7 times is better than running the 1-hour model 168 times as it significantly reduces the accumulation errors for 7-day forecasts. However, they acknowledged that training a model directly predicting the lead time beyond 24 hours is challenging with their current model. Meanwhile, Lam et al. employed a curriculum training schedule following pre-training to improve GraphCast's ability to make accurate forecasts for multiple steps \cite{lam2022graphcast}. Increasing autoregressive steps results in excessive memory and computational costs, thereby limiting the maximum feasible number of steps. Chen et al. \cite{chen2023fengwu} proposed a reply buffer mechanism to mimic the long-lead autoregressive forecasts with improved computational efficiency and reduced memory costs. The study by Lam et al. \cite{lam2022graphcast} revealed that GraphCast's performance decreases in short lead times and improves at longer lead times as the number of autoregressive steps increases. Thus, using a single model is insufficient for achieving the best performance for both short and long lead times. 

To conclude, significant progress have been achieved in ML-based weather forecasting, particularly in 10-day forecasts where the ML models have outperformed ECMWF HRES. However, further breakthroughs are necessary to address the issues related to iterative accumulated errors and enhance the accuracy of forecasts for longer lead times. The next significant goals are to achieve comparable performance to ECMWF ensemble, of which the ensemble mean (EM) often has greater skill than the deterministic forecasts for longer lead times, and to increase the forecast lead time beyond 10 days. The objective of this study is to reduce the accumulation error and generate ML-based weather forecasts for 15 days that have performance comparable to ECMWF EM. However, since a single model has been shown to be incapable of achieving optimal forecast performance across various forecast lead times, we propose a novel cascade ML model architecture for weather forecasting based on pre-trained models, each optimized for specific forecast time windows. As a result, we present FuXi \footnote{FuXi, (Chinese: 伏羲), the first of ancient China’s mythological emperors, is said to be the first weather forecaster of China. He created bagua (八卦), eight diagrams, which were used to explain the constitution of the universe and predict weather.} weather forecasting system that generates 15-day forecasts at the spatial resolution of $0.25^{\circ}$. FuXi is a cascade of models optimized for three sequential forecast time periods of 0-5 days, 5-10 days, and 10-15 days, respectively. The base FuXi model is an autoregressive model designed to efficiently extract complex features and learn relationships from a large volume of high-dimensional weather data. Specifically, 39 years of 6-hourly ECMWF ERA5 reanalysis data at a spatial resolution of $0.25^{\circ}$ are used for developing the FuXi system. The evaluation shows that FuXi significantly outperforms ECMWF HRES and achieves comparable performance to ECMWF EM for the first time. FuXi extends the skillful forecast lead time, as indicated by whether anomaly correlation coefficient (ACC) being greater than 0.6, to 10.5 and 14.5 days for $Z500$ and $T2M$, respectively. Moreover, ensemble forecasts provide greater values beyond EM by offering estimates of forecast uncertainty and enabling skillful predictions for longer lead times. Therefore, we developed the FuXi ensemble forecast by introducing perturbations to initial conditions and model parameters in order to generate ensemble forecasts. The evaluation based on the continuous ranked probability score (CRPS) demonstrates that the FuXi ensemble performs comparably to the ECMWF ensemble within a forecast lead time of 9 days for ${Z500}$, ${T850}$, mean sea-level pressure (${MSL}$), and ${T2M}$.



Overall, our contribution to this work can be summarized as follows:

\begin{itemize}

\item We propose a novel cascade ML model architecture for weather forecasting, which aims to reduce accumulation errors.

\item FuXi achieves comparable performance to ECMWF EM and extends the skillful forecast lead time (ACC \gt 0.6) to 10.5 and 14.5 days for $Z500$ and $T2M$, respectively.

\end{itemize}

\section{Dataset}
\subsection{ERA5}
ERA5 is the fifth generation of the ECMWF reanalysis dataset, providing hourly data of surface and upper-air parameters at a horizontal resolution of approximately 31 km and 137 model levels from January 1940 to the present day \cite{hersbach2020era5}. The dataset is generated by assimilating high-quality and abundant global observations using ECMWF's IFS model. Given its coverage and accuracy, the ERA5 data is widely regarded as the most comprehensive and accurate reanalysis archive. Therefore, we use the ERA5 reanalysis dataset as the ground truth for the model training. 

We use a subset of the ERA5 dataset spanning 39 years, which has a spatial resolution of $0.25^\circ$ ($721\times1440$ latitude-longitude grid points) and a temporal resolution of 6 hours. In this work, we focus on predicting 5 upper-air atmospheric variables at 13 pressure levels (50, 100, 150, 200, 250, 300, 400, 500, 600, 700, 850, 925, and 1000 hPa), and 5 surface variables. The 5 upper-air atmospheric variables are geopotential (${Z}$), temperature (${T}$), u component of wind (${U}$), v component of wind (${V}$), and relative humidity (${R}$). Additionally, 5 surface variables are ${T2M}$, 10-meter u wind component (${U10}$), 10-meter v wind component (${V10}$), ${MSL}$, and ${TP}$\footnote{A summary of variable definitions can be referred to in Table \ref{glossary}}. In total, 70 variables are predicted and evaluated.

Following previous studies in splitting the data into training, validation, and testing set \cite{pathak2022fourcastnet,lam2022graphcast}, the training set consists of 54020\footnote{54020 = 365 $\times$ 4 $\times$ 37, similarly, 2920 = 365 $\times$ 4 $\times$ 2, and 1460 = 365 $\times$ 4} samples spanning from 1979 to 2015. The validation set contains 2920 samples corresponding to the years 2016 and 2017, while out-of-sample testing is performed using 1460 samples from 2018.

\begin{table}
\centering
\caption{\label{glossary} A summary of variable definitions used in this paper.}
\begin{tabularx}{\textwidth}{cX}
\hline
\textbf{Variables} & \textbf{Definitions} \\
\hline
${C}$, ${H}$, ${W}$ & Channel dimensions, and spatial dimensions in latitude and longitude directions, respectively.  \\
${D}$ & A set containing all the forecast initialization times in the testing dataset. \\
${c}$, ${i}$, ${j}$, ${t_0}$, ${\tau}$ & Indices for variables, latitude coordinates, and longitude coordinates, as well as forecast initialization time and forecast lead time steps added to ${t_0}$, respectively. \\
${X^t}$, ${\hat{X}^t}$, ${M}$ & Ground truth and model predicted weather parameters at time step ${t}$, and climatological mean computed using ERA reanalysis data between 1993 and 2016. \\
${Z}$, ${R}$, ${T}$, ${U}$, ${V}$, ${TP}$, ${MSL}$ & They represent geopotential, relative humidity, temperature, u component of wind, v component of wind, 6-hourly total precipitation, and mean sea-level pressure, respectively. \\
\hline
\end{tabularx}
\end{table}

\subsection{HRES-fc0 and ENS-fc0 dataset}
In this study, we evaluate our model against the ERA5 reanalysis data. Besides, we also created two reference datasets, HRES-fc0 and ENS-fc0, which consist of the first time step of each HRES and ensemble control forecast, respectively. We use these datasets to assess the performance of ECMWF HRES and EM. This approach aligns with that used by Haiden et al. \cite{ECMWF2021} and Lam et al. \cite{lam2022graphcast} in evaluating ECMWF forecasts.

\section{Methodology} 

\subsection{Generating 15-day forecasts using FuXi} \label{fuxi.arch}

The FuXi model is an autoregressive model that leverages weather parameters $(X^{t-1}, X^t)$ from two previous time steps as input to forecast  weather parameters at the upcoming time step $(X^{t+1})$. ${t}$, ${t-1}$, and ${t+1}$ represent the current, the prior, and upcoming time steps, respectively. The time step considered in this model is 6 hours. By utilizing the model's outputs as inputs, the system can generate forecasts with different lead times.

Generating 15-day forecasts using a single FuXi model requires 60 iterative runs. Pure data-driven ML models, unlike physics-based NWP models, lack physical constraints, which can result in significantly growing errors and unrealistic predictions for long-term forecasts. Using an autoregressive, multi-step loss effectively minimizes accumulation error for long lead times \cite{lam2022graphcast}. This loss is similar to the cost function applied in the four-dimensional variational data assimilation (4D-Var) method, which aims to identify the initial weather conditions that optimally fit observations distributed over an assimilation time window. Although increasing the autoregressive steps leads to more accurate forecasts for longer lead times, it also results in less accurate results for shorter lead times. Besides, increasing autoregressive steps require more memory and computing resources for handling gradients during the training process, similar to increasing the assimilation time window of 4D-Var.

When making iterative forecasts, error accumulation is inevitable as lead times increase. Also, previous studies indicate that a single model can not perform optimally across all lead times. To optimize performance for both short and long lead times, we propose a cascade \cite{ho2021cascaded,Li2015} model architecture using pre-trained FuXi models, fine-tuned for optimal performance in specific 5-day forecast time windows. These windows are referred to as FuXi-Short (0-5 days), FuXi Medium (5-10 days), and FuXi-Long (10-15 days). As shown in Figure \ref{model}, FuXi-Short and FuXi Medium outputs from the 20th and 40th steps are used as inputs to FuXi-Medium and FuXi-Long, respectively. Unlike the greedy hierarchical temporal aggregation strategy employed in Pangu-Weather \cite{bi2022panguweather}, which utilizes 4 models with forecast lead times of 1 h, 3 h, 6 h, and 24 h to minimize the number of steps, the cascaded FuXi model does not suffer from temporal inconsistency. The cascaded FuXi model performs comparably to ECMWF EM in 15-day forecasts, as shown in Figure \ref{skills_vs_EM}.

\subsection{FuXi model architecture} 

\begin{figure}
    \centering
    \includegraphics[width=\linewidth]{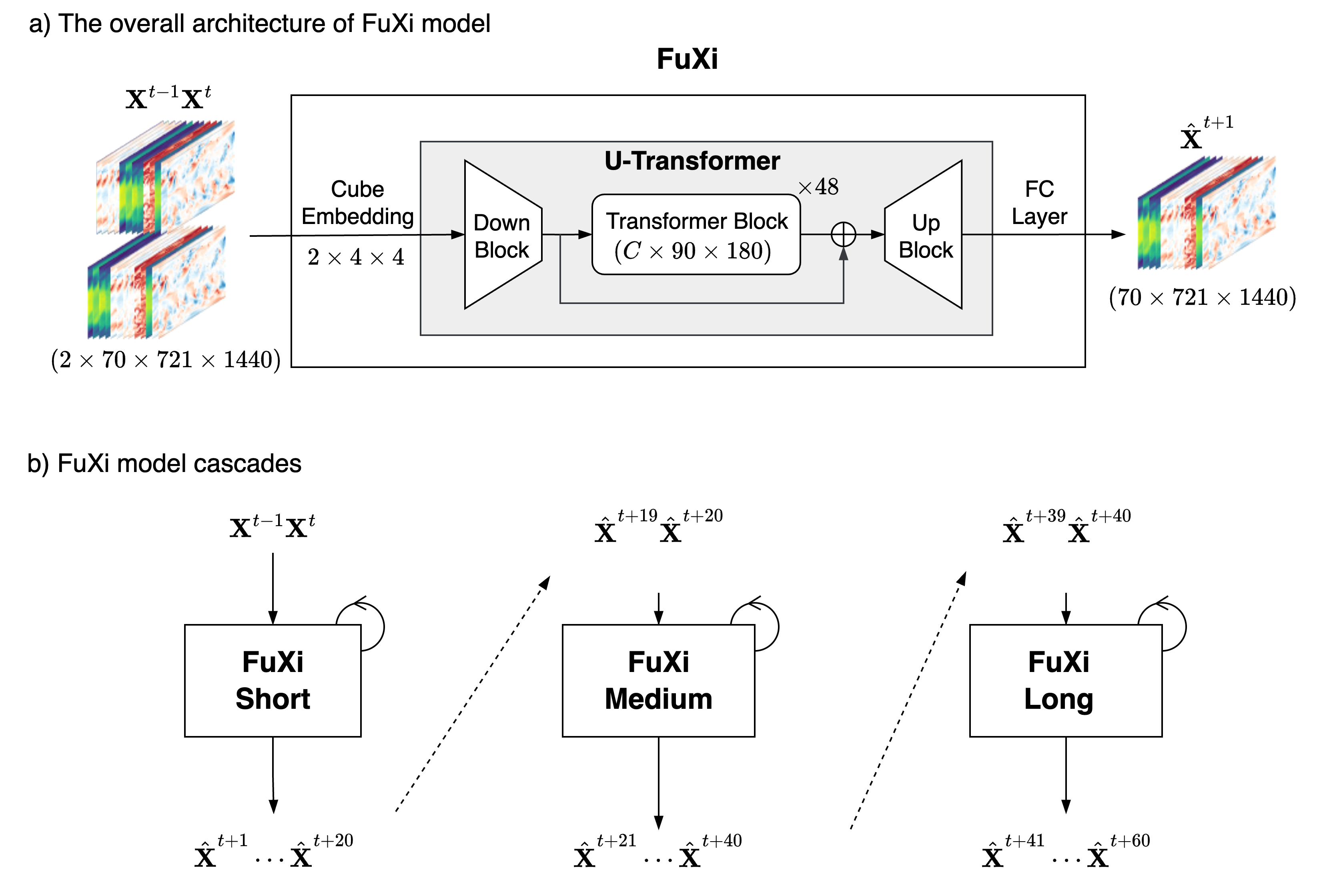}
    \caption{Overall architecture of FuXi model. a) The FuXi model consists of three components: cube embedding, U-Transformer, and fully connected (FC) layer; b) FuXi-Short, FuXi-Medium, and FuXi-Long models cascade and produce 15-day forecasts, with each model generating 5 days forecasts.}
    \label{model}    
\end{figure}

The model architecture of the base FuXi model consists of three main components, which are illustrated in Figure \ref{model}: cube embedding, U-Transformer, and a fully connected (FC) layer. The input data combines both upper-air and surface variables and creates a data cube with dimensions of $2\times70\times721\times1440$, where ${2}$, ${70}$, ${721}$, and ${1440}$ represent the two preceding time steps (${t-1}$ and ${t}$), the total number of input variables, latitude (${H}$) and longitude (${W}$) grid points, respectively. 

Firstly, the high-dimensional input data undergoes dimension reduction to $C\times180\times360$ through joint space-time cube embedding, where $C$ is the number of channels, and is set to be 1536). The primary purpose of cube embedding is to reduce the temporal and spatial dimensions of input data, making it less redundant. Subsequently, the U-Transformer processes the embedded data, and prediction follows using a simple FC layer. The output is initially reshaped to $70\times720\times1440$, then restored to the original input shape of $70\times721\times1440$ by bilinear interpolation. The following subsections provide details for each component in the base FuXi model.

\subsubsection{Cube embedding} 

To reduce the spatial and temporal dimensions of input and accelerate the training process, the space-time cube-embedding \cite{tong2022videomae} is applied. A similar approach, patch embedding, which divides an image into N $\times$ N patches with each patch being transformed into a feature vector, was used in the Pangu-Weather model \cite{bi2022panguweather}. The cube embedding applies a 3-dimensional (3D) convolution layer, with a kernel and stride of 2$\times$4$\times$4 (equivalent to $\frac{T}{2}\times\frac{H}{4}\times\frac{W}{4}$), and output channels numbering $C$. Following cube embedding, a layer normalization (LayerNorm) \cite{ba2016layer} is utilized to improve training stability. The result is a data cube with dimensions of $C\times180\times360$.

\subsubsection{U-Transformer} 

This subsection presents the design of the U-Transformer, which is visually represented in Figure \ref{model} through its schematic diagram.

Recently, the ViT \cite{ViT2021} and its variants have demonstrated remarkable performance in various computer vision tasks by using the multi-head self-attention, which enables the simultaneous processing of sequential input data. Nevertheless, global self-attention is infeasible for processing high-resolution inputs due to its quadratic computational and memory complexity with respect to the input size. Swin Transformer was proposed as a solution \cite{liu2021swin} to improve computational efficiency by limiting computation of self-attention only within the non-overlapping local windows. Besides, the shifted-window mechanism allows for cross-connections between windows. As a result, the Swin Transformer has shown superior performance on various benchmarks and is frequently used as a backbone architecture in many vision tasks. Additionally, many researchers have developed ML-based weather forecasting models using Swin Transformer blocks \cite{hu2022swinvrnn,bi2022panguweather,pathak2022fourcastnet,swinRDM2023}. 

However, training and applying a large-scale Swin Transformer model for high-resolution inputs reveals several issues, including training instability. To address these issues, Swin Transformer V2 \cite{liu2022swin} was proposed, which upgrades the original Swin-Transformer (V1) \cite{liu2021swin} by using the residual post-normalization instead of pre-normalization \footnote{The LN layer is moved from the beginning of each residual unit to the end, producing much milder activation values.}, scaled cosine attention instead of the original dot product self-attention \footnote{This makes the computation irrelevant to amplitudes of block inputs so that attention values are less likely to fall into extremes.}, and log-spaced coordinates instead of the previous linear-spaced coordinates. As a result, Swin Transformer V2 has 3 billion parameters and advances state-of-the-art performance on multiple vision task benchmarks.

The U-Transformer is constructed using 48 repeated Swin Transformer V2 blocks and calculates the scaled cosine attention as follows:
\begin{equation} \label{attn}
     \text{Attention}\textbf{(Q, K, V)} = (\cos\textbf{(Q, K)}/ \tau + \textbf{B}) \textbf{V}
\end{equation}
where $\textbf{B}$ represents the relative position bias and $\tau$ is a learnable scalar, which is not shared across heads and layers. The cosine function is naturally normalized, which leads to smaller attention values.

The U-Transformer, as the name implies, also includes a downsampling and upsampling block from the U-Net model \cite{ronneberger2015u}. The downsampling block, referred to as the Down Block in Figure \ref{model}, reduces the data dimension to $C\times90\times180$, thereby minimizing computational and memory requirements for self-attention calculation. The Down Block consists of a $3\times3$ 2-dimensional (2D) convolution layer with a stride of 2, and a residual \cite{he2015deep} block that has two $3\times3$ convolution layers followed by a group normalization (GN) layer \cite{wu2018group} and a sigmoid-weighted linear unit (SiLU) activation \cite{elfwing2017sigmoidweighted,ramachandran2017searching}. The SiLU activation is calculated by multiplying the sigmoid function with its input ($\sigma$({x})$\times${x}).
The upsampling block, known as Up Block in Figure \ref{model}, has the same residual block as used in the Down Block, along with a 2D transposed convolution \cite{Zeiler2010} with a kernel of 2 and a stride of 2. The Up Block scales the data size back up to $C\times180\times360$. Furthermore, a skip connection is included that concatenates the outputs from the Down Block with those of the transformer blocks before being fed into the Up Block.

\subsection{FuXi model training} \label{fuxi.cascade}

This section outlines the training process for FuXi models. The training procedure involves two steps: pre-training and fine-tuning, similar to the approach used for training GraphCast \cite{lam2022graphcast}.

\subsubsection{One-step Pre-training}

The pre-training step involves supervised training and optimizing the FuXi model to predict a single time step using the training dataset. The loss function used is the latitude-weighted ${L1}$ loss, which is defined as follows:

\begin{equation}
    L1 = \frac{1}{C\times H \times W} \sum_{c=1}^{C}\sum_{i=1}^H\sum_{j=1}^{W} a_i \mid \hat{X}^{t+1}_{c,i,j} - X^{t+1}_{c,i,j} \mid
\end{equation}
where $C$, $H$, and $W$ are the number of channels and the number of grid points in latitude and longitude direction, respectively. ${c}$, ${i}$, and ${j}$ are the indices for variables, latitude and longitude coordinates, respectively. ${\hat{X}^{t+1}_{c,i,j}}$ and ${X^{t+1}_{c,i,j}}$ are predicted and ground truth for some variable and locations (latitude and longitude coordinates) at time step of ${t+1}$. $a_i$ represents the weight at latitude ${i}$ and the value of $a_i$ decreases as latitude increases. The ${L1}$ loss is averaged over all the grid points and variables.

The FuXi model is developed using the Pytorch framework \citep{Paszke2017}. Pre-training of the model requires approximately 30 hours on a cluster of 8 Nvidia A100 GPUs. The model is trained with 40000 iterations using a batch size of 1 on each GPU. The AdamW \cite{kingma2017adam,loshchilov2017decoupled} optimizer is used with parameters ${\beta_{1}}$=0.9 and ${\beta_{2}}$=0.95, an initial learning rate of 2.5$\times$10$^{-4}$, and a weight decay coefficient of 0.1. Scheduled DropPath \cite{larsson2017fractalnet} with a dropping ratio of 0.2 is employed to prevent overfitting. In addition, Fully-Sharded Data Parallel (FSDP) \cite{zhao2023pytorch}, bfloat16 floating point precision, and gradient check-pointing \cite{chen2016training} are applied to reduce memory costs during model training. 

\subsubsection{Fine-tuning cascaded models}
After pre-training, the base FuXi model is first fine-tuned for optimal performance for 6-hourly forecasts spanning from 0 to 5 days (0-20 time steps). This fine-tuning process is performed using an autoregressive training regime and curriculum training schedule to increase the number of autoregressive steps from 2 to 12, following the fine-tuning approach of the GraphCast model \cite{lam2022graphcast}. This fine-tuned model is referred to as FuXi-Short in Figure \ref{model}. With weights from FuXi-Short, the FuXi-Medium model is initialized and then fine-tuned for optimal forecast performance for 5 to 10 days (21-40 time steps). Implementing the online inference of FuXi-Short to get output at the 20th time step (5th day), which is required for input to the FuXi-Medium model during its fine-tuning process, is inappropriate due to significant memory consumption and the slowdown of the fine-tuning process for FuXi-Medium. To address this issue, the results of FuXi-Short for six years of data (2012-2017) are cached on a hard disk beforehand. The same procedure for fine-tuning FuXi-Medium is repeated for the fine-tuning of FuXi-Long, optimized for generating forecasts of 10-15 days. Finally, FuXi-Short, FuXi-Medium, and FuXi-Long are cascaded to produce the complete 15-day forecasts. As detailed in Appendix \ref{Effectiveness_cascade}, cascade helps to reduce accumulation errors and improve forecast performance for longer lead times.

During the fine-tuning process, the model was trained using a constant learning rate of 1$\times$10$^{-7}$. It takes approximately two days to fine-tune each of the cascaded FuXi models on a cluster of 8 Nvidia A100 GPUs.

\subsection{FuXi ensemble forecast}\label{FuXi_ENS}
Weather forecasting is an inherently uncertain due to the chaotic nature of the weather system \cite{Lorenz1963}. To address this uncertainty, ensemble forecasting is necessary, particularly for longer lead times. Additionally, since ML models can generate forecasts at significantly lower computational costs compared to conventional NWP models, we generated a 50-member ensemble forecast using the FuXi model. Following the approach used by ECMWF for ensemble runs, which involves perturbing both initial conditions and model physics \cite{buizza1999stochastic,leutbecher2008ensemble}, we incorporated random Perlin noise \cite{bi2022panguweather} into the initial conditions and implemented the Monte Carlo dropout (MC dropout, dropout rate is 0.2) \cite{gal2016dropout} to perturb the model parameters. More specifically, each of the 49 perturbations contains 4 octaves of Perlin noise, a scaling factor of 0.5, and the number of periods of noise to generate along each axis (channel, latitude, and longitude) being 1, 6 and 6, respectively.

\subsection{Evaluation method}\label{evaluation_method}

We follow \cite{rasp2020weatherbench} to evaluate forecast performance using latitude-weighted root mean square error (RMSE) and ACC, which are calculated as follows:

\begin{equation}
    RMSE(c, \tau) =\frac{1}{\mid D \mid}\sum_{t_0 \in D} \sqrt{\frac{1}{H \times W} \sum_{i=1}^H\sum_{j=1}^{W} a_i {( \hat{X}^{t_0 +\tau}_{c,i,j} - X^{t_0 +\tau}_{c,i,j} )}^{2}}
\end{equation}

\begin{equation}
    ACC(c, \tau) = \frac{1}{\mid D \mid}\sum_{t_0 \in D} \frac{\sum_{i, j} a_i (\hat{X}^{t_0 +\tau}_{c,i,j} - M^{t_0 +\tau}_{c,i,j}) (\hat{X}^{t_0 +\tau}_{c,i,j} - M^{t_0 +\tau}_{c,i,j})} {\sqrt{ \sum_{i, j} a_i (\hat{X}^{t_0 +\tau}_{c,i,j} - M^{t_0 +\tau}_{c,i,j})^2 \sum_{i, j} a_i(\hat{X}^{t_0 +\tau}_{c,i,j} - M^{t_0 +\tau}_{c,i,j})^2}}
\end{equation}
where ${t_0}$ is the forecast initialization time in the testing set ${D}$, and ${\tau}$ is the forecast lead time steps added to ${t_0}$. ${M}$ represents the climatological mean calculated using ERA5 reanalysis data between 1993 and 2016. Additionally, to improve the discrimination of the forecast performance among models with small differences, we use the normalized RMSE difference between model ${A}$ and baseline ${B}$ calculated as \((RMSE_A-RMSE_B)/RMSE_B\), and the normalized ACC difference represented by \((ACC_A-ACC_B)/(1-ACC_B)\). Negative values in normalized RMSE difference and positive values in normalized ACC difference indicate that model ${A}$ performs better than the baseline model ${B}$.

To evaluate the performance of ECMWF HRES and EM, the verification method implemented by ECMWF \cite{ECMWF2021} is used where the model analysis, namely HRES-fc0 and ENS-fc0, serve as the ground truth for HRES and EM, respectively.

In addition, we assess the quality of ensemble forecasts by calculating two metrics: the CRPS \cite{Hersbach2000,Sloughter2010} and the spread-skill ratio (SSR). The CRPS is computed using the following equation:
\begin{equation}
CRPS = \int_{-\infty}^{\infty} [F(\hat{X}^{t_0 +\tau}_{c,i,j})-\mathcal{H}(X^{t_0 +\tau}_{c,i,j} \le z)] \,dz \
\end{equation}
where $F$ represents the cumulative distribution function (CDF) of the forecasted variable ($\hat{X}^{t_0 +\tau}_{c,i,j}$), and $\mathcal{H}$ is an indicator function. The indicator function equals 1 if the statement $X^{t_0 +\tau}_{c,i,j} \le z$ is true; otherwise takes the value of 0 \cite{Wilks2011}. For deterministic forecasts, the CRPS reduces to the mean absolute error (MAE) \cite{Hersbach2000}. The xskillscore Python package is used to calculate the CRPS metric. And we assume that the distribution of ensemble members follows a Gaussian distributions, and the CRPS is computed based on the ensemble mean and the ensemble variance. On the other hand, the SSR measures the consistency between the spread of the ensemble and the RMSE of the EM. The ensemble spread is defined as:
\begin{equation}
    Spread(c, \tau) =\frac{1}{\mid D \mid}\sum_{t_0 \in D} \sqrt{\frac{1}{H \times W} \sum_{i=1}^H\sum_{j=1}^{W} a_i var( \hat{X}^{t_0 +\tau}_{c,i,j} )}
\end{equation}
where $var( \hat{X}^{t_0 +\tau}_{c,i,j} )$ denotes the variance within the ensemble dimension. A reliable ensemble is indicated by a SSR of one \cite{Fortin2014}. Lower values suggest an underdispersive ensemble forecast, while higher values indicate overdispersion.

\section{Results}

For evaluating FuXi's performance, the study uses the 2018 data and selects two daily initialization times (00:00 UTC and 12:00 UTC) to produce 6-hourly forecasts for 15 days.

\subsection{Deterministic forecast metrics comparison}

\begin{figure}
    \centering
    \includegraphics[width=\linewidth]{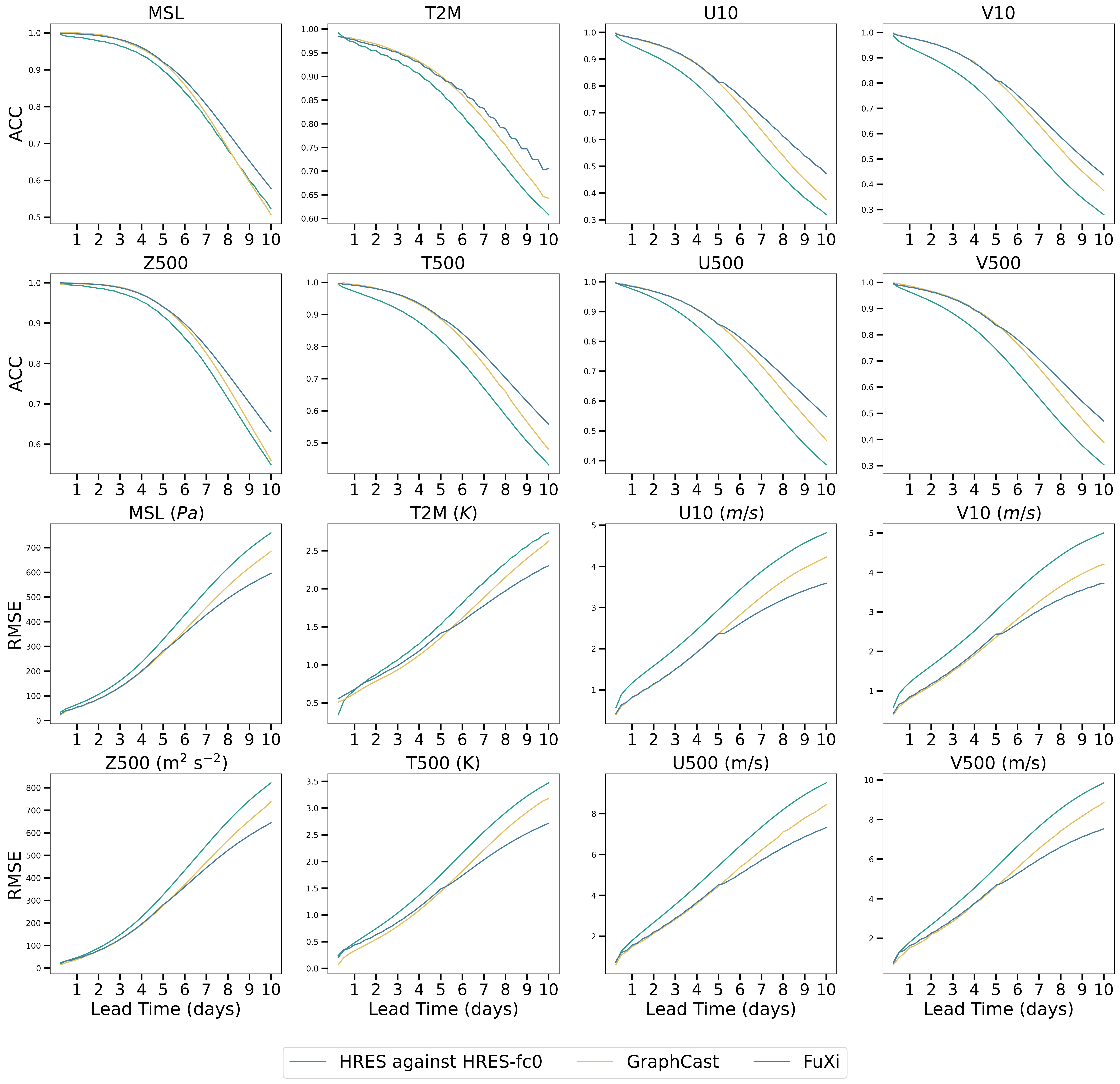}
    \caption{Comparison of the globally-averaged latitude-weighted ACC (first and second rows) and RMSE (third and fourth rows) of the HRES (dark green lines), GraphCast (organge lines), and FuXi (light blue lines) for 4 surface variables, such as ${MSL}$, ${T2M}$, ${U10}$, and ${V10}$, and 4 upper-air variables at the pressure level of 500 hPa, including ${Z500}$, ${T500}$, ${U500}$, and ${V500}$, using testing data from 2018. FuXi and GraphCast are evaluated against the ERA5 reanalysis dataset, and ECMWF HRES is evaluated against HRES-fc0.}
    \label{skills_vs_HRES}
\end{figure}

This subsection compares the forecast performance of FuXi, ECMWF HRES, GrahpCast (the state-of-the-art ML-based weather forecast), ECMWF EM, and FuXi EM on deterministic metrics. Figure \ref{skills_vs_HRES} shows the time series of the globally-averaged latitude-weighted ACC and RMSE of FuXi, ECMWF HRES, and GraphCast for 4 surface variables (${MSL}$, ${T2M}$, ${U10}$, and ${V10}$) and 4 upper-air variables (${Z500}$, ${T500}$, ${U500}$, and ${V500}$) at 500 hPa pressure level. The figure illustrates that both FuXi and GraphCast significantly outperform ECMWF HRES. FuXi and GraphCast have comparable performance within forecasts of 7 days, beyond which FuXi shows superior performance, with the lowest values of RMSE and the highest values of ACC across all the variables and forecast lead times. Moreover, FuXi's superior performance becomes increasingly significant as lead times increase. Using an ACC value of 0.6 as the threshold to measure a skillful weather forecast, we find that FuXi extends the skilful forecast lead time compared to ECMWF HRES, especially pushing the lead time of ${Z500}$ and ${T2M}$ from 9.25 and 10 days to 10.5 and 14.5 days (see Figure \ref{skillful_lead} for comparison of skillful forecast lead time), respectively.


\FloatBarrier

\begin{figure}
    \centering
    \includegraphics[width=\linewidth]{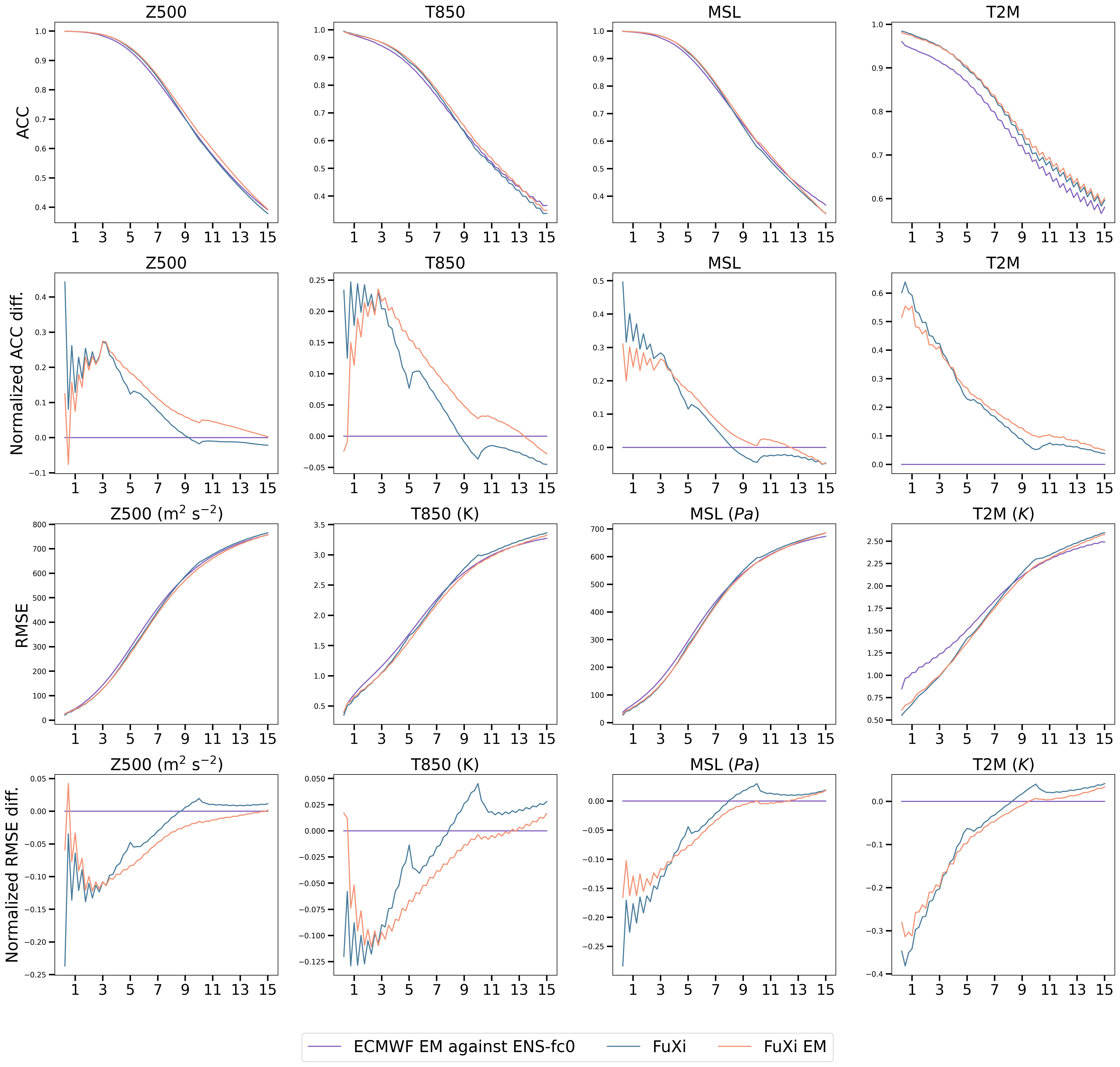}
    \caption{Comparison of the globally-averaged latitude-weighted ACC (first row) and RMSE (third row) as well as normalized ACC (second row) and RMSE difference (fourth row) of ECMWF EM (light purple lines), FuXi (light blue lines), and FuXi EM (light red lines) for 2 upper-air variables, including ${Z500}$ (first column) and ${T850}$ (second column), and 2 surface variables, such as ${MSL}$ (third column) and ${T2M}$ (fourth column), in 15-day forecasts using testing data from 2018. FuXi and FuXi EM is evaluated against the ERA5 reanalysis dataset, and ECMWF ensemble is evaluated against ENS-fc0.}
    \label{skills_vs_EM}
\end{figure}

Figure \ref{skills_vs_EM} shows the time series of the globally-averaged latitude-weighted ACC and RMSE of ECMWF EM, FuXi, and FuXi EM, as well as the corresponding normalized differences in ACC and RMSE for 4 variables. The 4 variables include 2 upper-air variables (${Z500}$ and ${T850}$) and 2 surface variables (i.e., ${MSL}$ and ${T2M}$). Many combinations of variables and pressure levels are not included in the comparisons as they are unavailable from the ECMWF server. The normalized differences in ACC and RMSE are computed using ECMWF EM as the reference, as shown in the 2nd and 4th rows of Figure \ref{skills_vs_EM}. FuXi superior performance to ECMWF EM in 0-9 day forecasts, with positive values in the normalized ACC difference and negative values in normalized RMSE difference. However, for forecasts beyond 9 days, FuXi shows slightly poorer performance compared to ECMWF EM. Overall, FuXi shows comparable performance to ECMWF EM in 15-day forecasts, with higher ACC and lower RMSE than ECMWF EM on 67.92\% and 53.75\% of the 240 combinations of variables, levels, and lead times in the testing set, which includes 2 surface variable and 2 upper-air variables over 15 days, with 4 steps each day. The higher percentage of ACC could potentially be attributed to the fact that the climatological mean used in the computation of the ACC is based on ERA5 data, which serves as the ground truth for training FuXi. FuXi EM is slightly inferior to the Fuxi deterministic forecast within short lead times for all variables shown in Figure \ref{skills_vs_EM}. However, it performs better after the lead time surpasses 3 days, which aligns with Pangu-Weather and FourCastNet.

\FloatBarrier

\begin{figure}
    \centering
    \includegraphics[width=\linewidth]{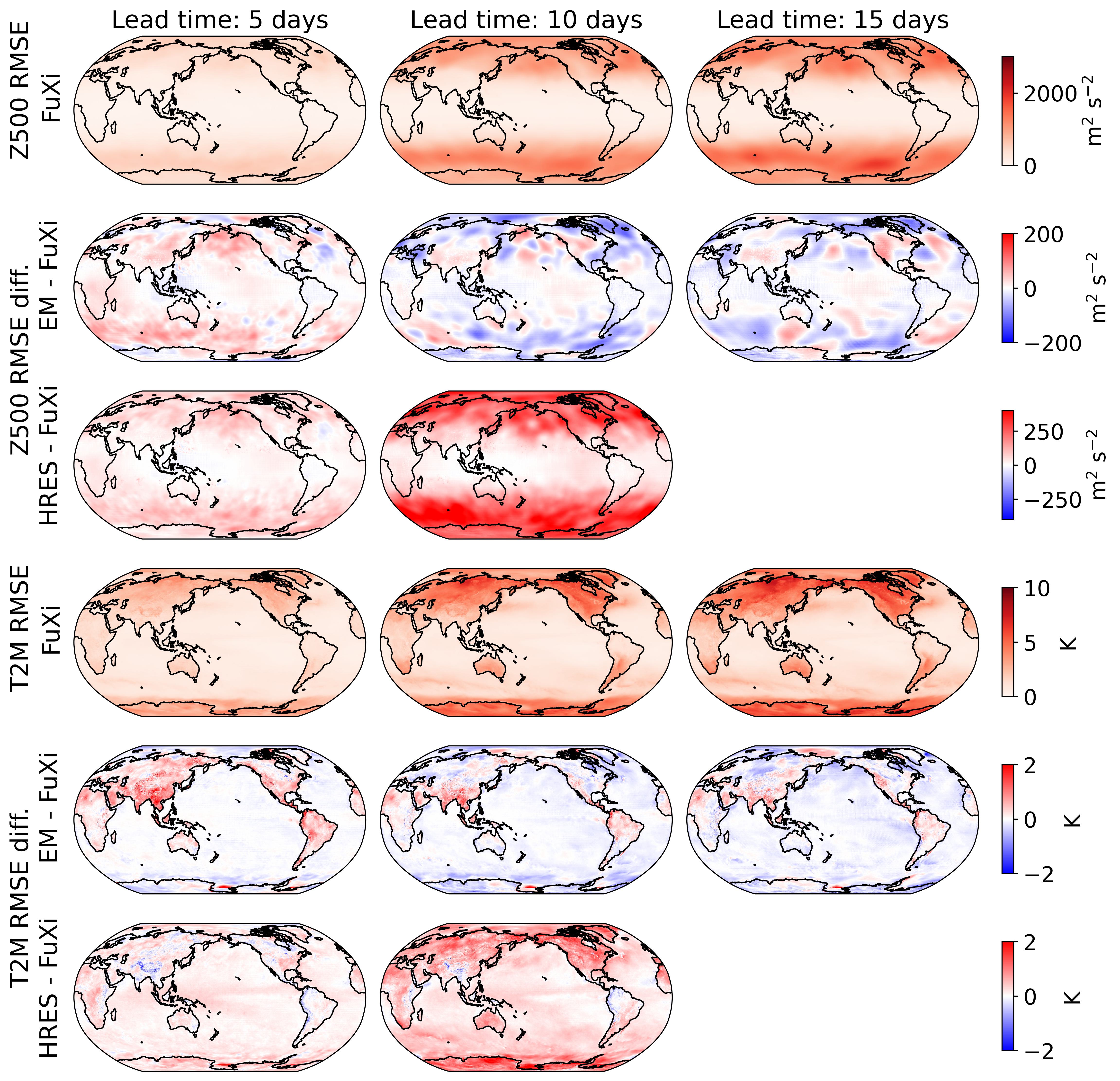}
    \caption{Spatial map of average RMSE (not latitude-weighted) of FuXi (first and fourth rows), the difference in RMSE between ECMWF EM (second and fifth rows) and FuXi, and the difference in RMSE between ECMWF HRES (third and sixth rows) and FuXi for ${Z500}$ (first to third rows) and ${T2M}$ (fourth to sixth rows) at forecast lead times of 5 days (first column), 10 days (second column), and 15 days (third column), using the 2018 testing data.}
    \label{spatial_error}
\end{figure}

Figure \ref{spatial_error} illustrates the spatial distributions of the average RMSE of FuXi, the RMSE difference between ECMWF HRES and FuXi, and the RMSE difference between ECMWF EM and FuXi for forecasts of ${Z500}$ and ${T2M}$ at lead times of 5 days, 10 days, and 15 days, respectively. All forecasts in the testing data from 2018 were averaged to produce the data. The RMSE difference is represented by red, blue, and white patterns indicating whether ECMWF HRES or ECMWF EM performs worse than, better than, or equally compared to FuXi. Overall, all three forecasts have similar spatial error distributions, with the RMSE difference values much lower than the RMSE values. The highest RMSE values appear at high latitudes, while relatively small values are found in middle and low latitudes. The values of RMSE are higher over the land than over the ocean. The RMSE difference between ECMWF HRES and FuXi shows that FuXi outperforms ECMWF HRES in most grid points, as shown by the predominance of red color. In contrast, ECMWF EM shows comparable performance to FuXi in most areas, as indicated by the predominantly white color.

\FloatBarrier

\subsection{Ensemble forecast metrics comparison}
\begin{figure}
    \centering
    \includegraphics[width=\linewidth]{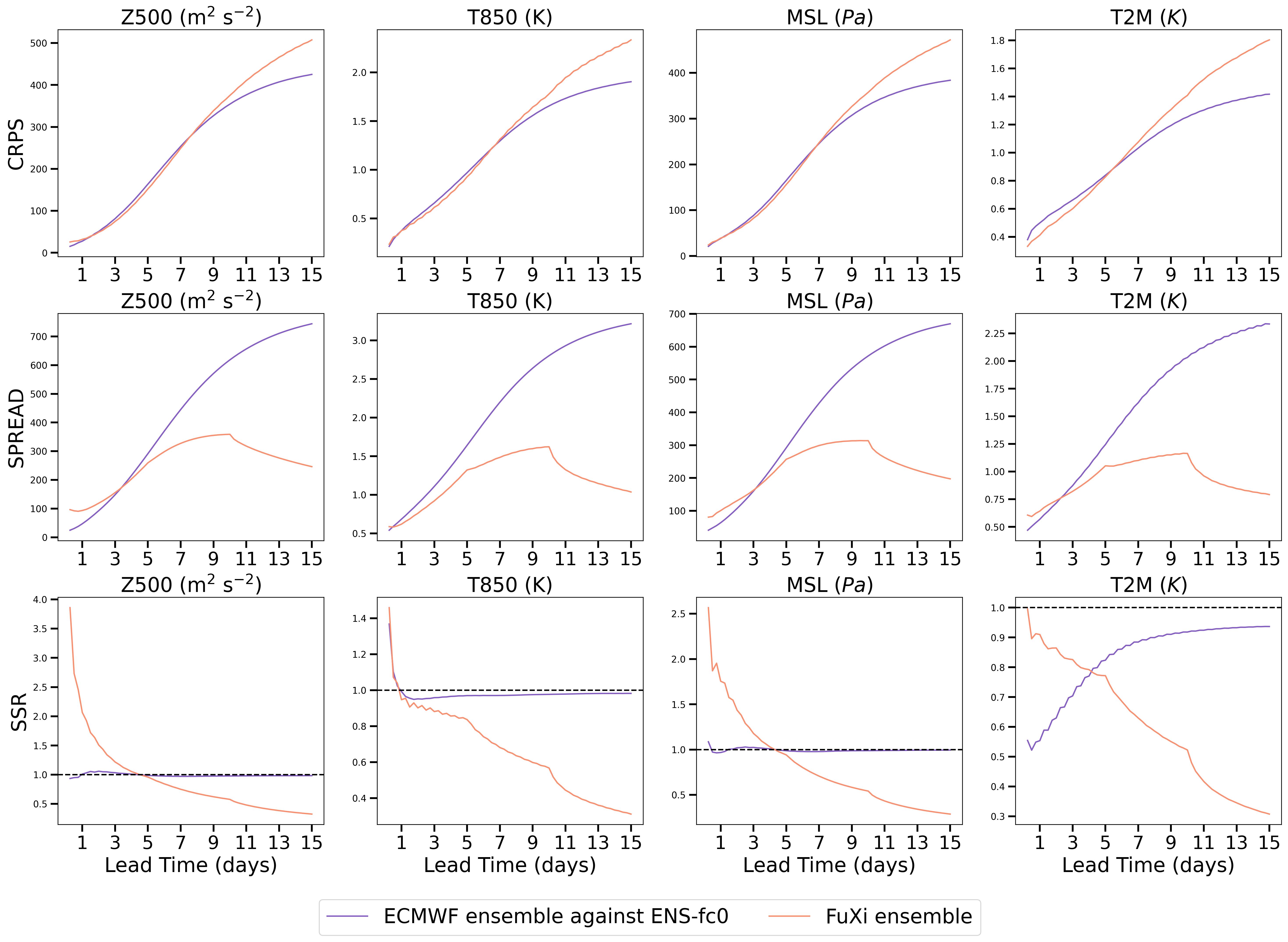}
    \caption{Comparison of the CRPS (first row), the globally-averaged latitude-weighted Spread (second row) and SSR (third row) of the ECMWF ensemble (light purple lines) and the FuXi ensemble (light red lines) for 2 upper-air variables, including ${Z500}$ (first column) and ${T850}$ (second column), and 2 surface variables, such as ${MSL}$ (third column) and ${T2M}$ (fourth column), in 15-day forecasts using testing data from 2018. The FuXi ensemble is evaluated against the ERA5 reanalysis dataset, and the ECMWF ensemble is evaluated against ENS-fc0.}
    \label{skills_vs_ensemble}
\end{figure}

Compared to deterministic forecasts, ensemble forecasts have several advantages. They provide a more accurate EM than deterministic forecast in terms of deterministic metrics and also represent forecast uncertainty through the ensemble spread. This subsection focuses on comparing ensemble evaluation metrics between the FuXi ensemble and the ECMWF ensemble. Figure \ref{skills_vs_ensemble} illustrates the time series of the CRPS, globally-averaged latitude-weighted Spread and SSR for the same 4 variables as shown in Figure \ref{skills_vs_EM}. The CRPS values for the FuXi ensemble are comparable to those of the ECMWF ensemble and slightly smaller before 9 days. However, beyond 9 days, the FuXi ensemble demonstrates inferior CRPS compared to the ECMWF ensemble. The SSR values for the FuXi ensemble are significantly higher than 1 for the 3 variables such as ${Z500}$, ${T850}$, and ${MSL}$ in early lead times, indicating overdispersion. These values then decrease dramatically with increasing lead times, and becomes lower than 1, indicating an underdispersive ensemble. Meanwhile, the SSR of the ECMWF ensemble are very close to 1, except for ${T2M}$. Both the FuXi ensemble and the ECMWF ensemble show underdispersion for ${T2}$ as their SSR values remain smaller than 1 throughout the 15-day forecast. While the ensemble spread of the ECMWF ensemble grows as the forecast lead time increases, the ensemble spread of the FuXi ensemble initially increases as the lead time increases, then decreases after 9 days. One plausible explanation is that the initial conditions are perturbed by the addition of Perlin noise, which is random and independent of the background flow. As a result, only a small fraction of the perturbations remains after 9 days of model integration, causing the ensemble spread to decrease.

\FloatBarrier

\section{Conclusion and Future Work} 

It has been challenging for data-driven methods to compete with conventional physics-based numerical weather prediction models in weather forecasting due to the difficulty in reducing accumulation error. Recently, ML-based weather forecasting systems have witnessed significant breakthroughs, outperforming ECMWF HRES in 10-day forecasts with a temporal resolution of 6 hours and a spatial resolution of $0.25^{\circ}$ \cite{bi2022panguweather,lam2022graphcast}. However, employing a single model proves insufficient to obtain optimal performance across various lead times. In order to generate skillful weather forecasts for longer lead times, such as 15 days, we first develop a powerful base ML model architecture, FuXi model. The FuXi model is based on the U-Transformer and has the capability to efficiently learn complex relationships from vast amounts of high-dimensional weather data. Moreover, we propose a novel cascade ML model architecture for weather forecasting that utilizes three pre-trained FuXi models. Each model is fine-tuned for optimal forecast performance for one of the forecast time windows: 0-5 days, 5-10 days, and 10-15 days. These models are then cascaded to generate comprehensive 15-day forecasts. By implementing the aforementioned methodologies, we created FuXi, an ML-based weather forecasting system that, for the first time, performs comparably to ECMWF EM in 15-day forecasts with a temporal and spatial resolution of 6 hours and $0.25^{\circ}$. Additionally, the FuXi ensemble forecast exhibits promising potential, with a comparable CRPS to ECMWF ensemble within 9 days for ${Z500}$, ${T850}$, ${MSL}$, and ${T2M}$.


In this study, we incorporate Perlin noise into the initial conditions to generate ensemble forecasts. The Perlin noise is random and independent of the background flow. Previous studies \cite{MAGNUSSON2009,Du2021} have shown that flow-independent initial perturbations decay over time during the model integration. Consequently, to ensure an adequate ensemble spread in the medium range, we will investigate flow-dependent methods for initial condition perturbations in order to maintain a reasonable spread throughout longer lead times for the FuXi ensemble.

Furthermore, we plan to explore the potential of utilizing the cascade ML model architecture for sub-seasonal forecasting. This will involve fine-tuning additional models for forecast lead times ranging from 14 to 28 days. Sub-seasonal forecasting remains a challenge and is considered as a "predictability desert" \cite{Vitart2012SubseasonalTS}. Unlike medium-range weather forecasting, which can utilize deterministic methods, ensemble forecasts are necessary for sub-seasonal forecasting. In addition, research has identified various processes in the atmosphere, ocean, and land that contribute to sub-seasonal predictability, such as the Madden-Julian Oscillation (MJO), soil moisture, snow cover, Stratosphere-troposphere interaction, and ocean conditions \cite{Robertson2018}. Therefore, more research is needed to develop an ML-based sub-seasonal forecasting system.

In addition, one limitation of current ML-based weather forecasting methods is that they are not yet completely end-to-end. They still rely on analysis data generated by conventional NWP models for initial conditions. Thus, we aim to develop a data-driven data assimilation method that uses observation data to generate initial conditions for ML-based weather forecasting systems. Looking to the future, we aim to build a truly end-to-end, systematically unbiased, and computationally efficient ML-based weather forecasting system.

\section*{Data Availability Statement}
We downloaded a subset of the ERA5 dataset from the official website of Copernicus Climate Data (CDS) at https://cds.climate.copernicus.eu/. The ECMWF HRES forecasts are available at \url{https://apps.ecmwf.int/archive-catalogue/?type=fc&class=od&stream=oper&expver=1} and ECMWF EM are available at \url{https://apps.ecmwf.int/archive-catalogue/?type=em&class=od&stream=enfo&expver=1}. The preprocessed sample data used for running FuXi models in this work are available in a Google Drive folder (\url{https://drive.google.com/drive/folders/1NhrcpkWS6MHzEs3i_lsIaZsADjBrICYV}) \cite{code2023}.

\section*{Code Availability Statement}
We used the code base of Swin transformer V2 as the backbone architecture, available at https://github.com/microsoft/Swin-Transformer. The source code used for training and running FuXi models in this work is available in a Google Drive folder (\url{https://drive.google.com/drive/folders/1NhrcpkWS6MHzEs3i_lsIaZsADjBrICYV}) \cite{code2023}. The aforementioned Google Drive folder contains the FuXi model, code, and sample input data, which can be accessed by individuals with the provided link. As the FuXi model and code are essential resources for this study, we have implemented password protection for the Google Drive folder link through a Google Form. To obtain the link to the Google Drive folder from the Zenodo link, users are required to complete the designated Google Form (\url{https://docs.google.com/forms/d/e/1FAIpQLSfjwZLf6PmxRvRhIPMQ1WRLJ98iLxOq_0dXb87N8CFNPyYAGg/viewform?usp=sharing}). 

The xskillscore Python package can be accessed from \url{https://github.com/xarray-contrib/xskillscore/}. The implementation of Perlin noise is based on publicly available from the GitHub repository: \url{https:// github.com/pvigier/perlin-numpy}.

\section*{Acknowledgements}

We appreciate the researchers at ECMWF for their efforts in collecting, archiving, disseminating, and maintaining the ERA5 reanalysis dataset, HRES, and ensemble, without which this study would not have been feasible. 

\section*{Competing interests}
The authors declare no competing interests.

\section*{Author Contributions}
H.L., Y.Q., and L.C. designed the project. H.L. and Y.Q. managed and oversaw the project. L.C. performed the model training and evaluation, and H.L. improved the model design. L.C., X.Z., and H.L. wrote and revised the manuscript. F.Z, Y.X., and Y.C. established the model training environment.

\noindent


\bibliography{refs}


\end{CJK*}

\clearpage

\appendix

\renewcommand\thefigure{\thesection.\arabic{figure}}    

\setcounter{figure}{0}    

\section*{Appendix}

\section{Effectiveness of the cascade model architecture}\label{Effectiveness_cascade}

\begin{figure}[h]
    \centering
    \includegraphics[width=\linewidth]{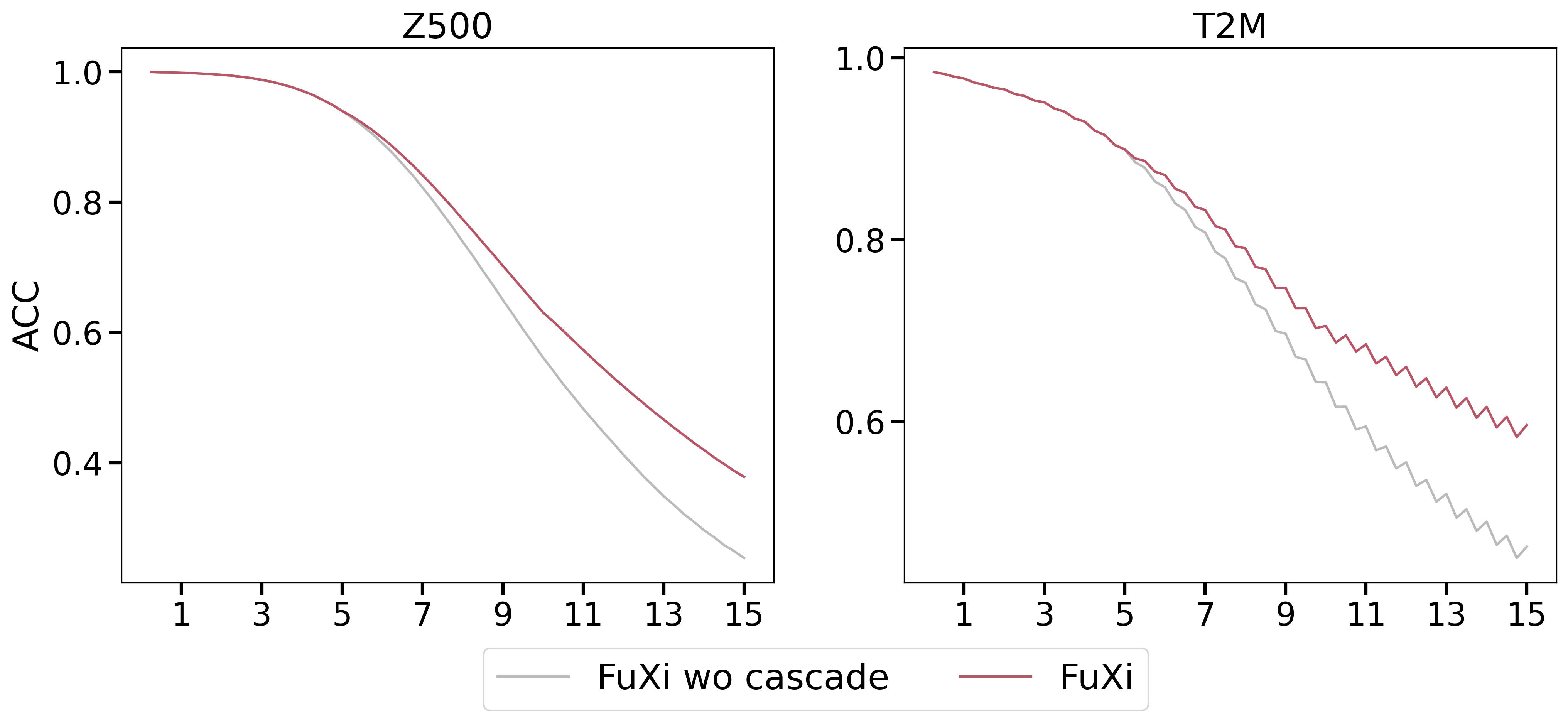}
    \caption{Comparison of the globally-averaged latitude-weighted ACC of the FuXi system (red lines) and FuXi without cascade (gray lines) using testing data from 2018.}
    \label{metric_cascade}    
\end{figure}

This section discusses the effect of using the cascade ML model architecture to reduce accumulation errors in weather forecasting. We use a single FuXi base model (FuXi-short) without the cascade, to generate 15-day forecasts. Then we evaluate its performance in comparison with the original FuXi model. Figure \ref{metric_cascade} shows the comparable performance for forecasts of ${Z500}$ and ${T2M}$ between the single base FuXi model and FuXi over lead times ranging from 0 to 7 days. However, the performance of the single FuXi base model deteriorates significantly as lead times increase, primarily due to accumulation error.

\section{Skillful forecast lead time comparison}\label{Effectiveness_cascade}

\begin{figure}[h]
    \centering
    \includegraphics[width=\linewidth]{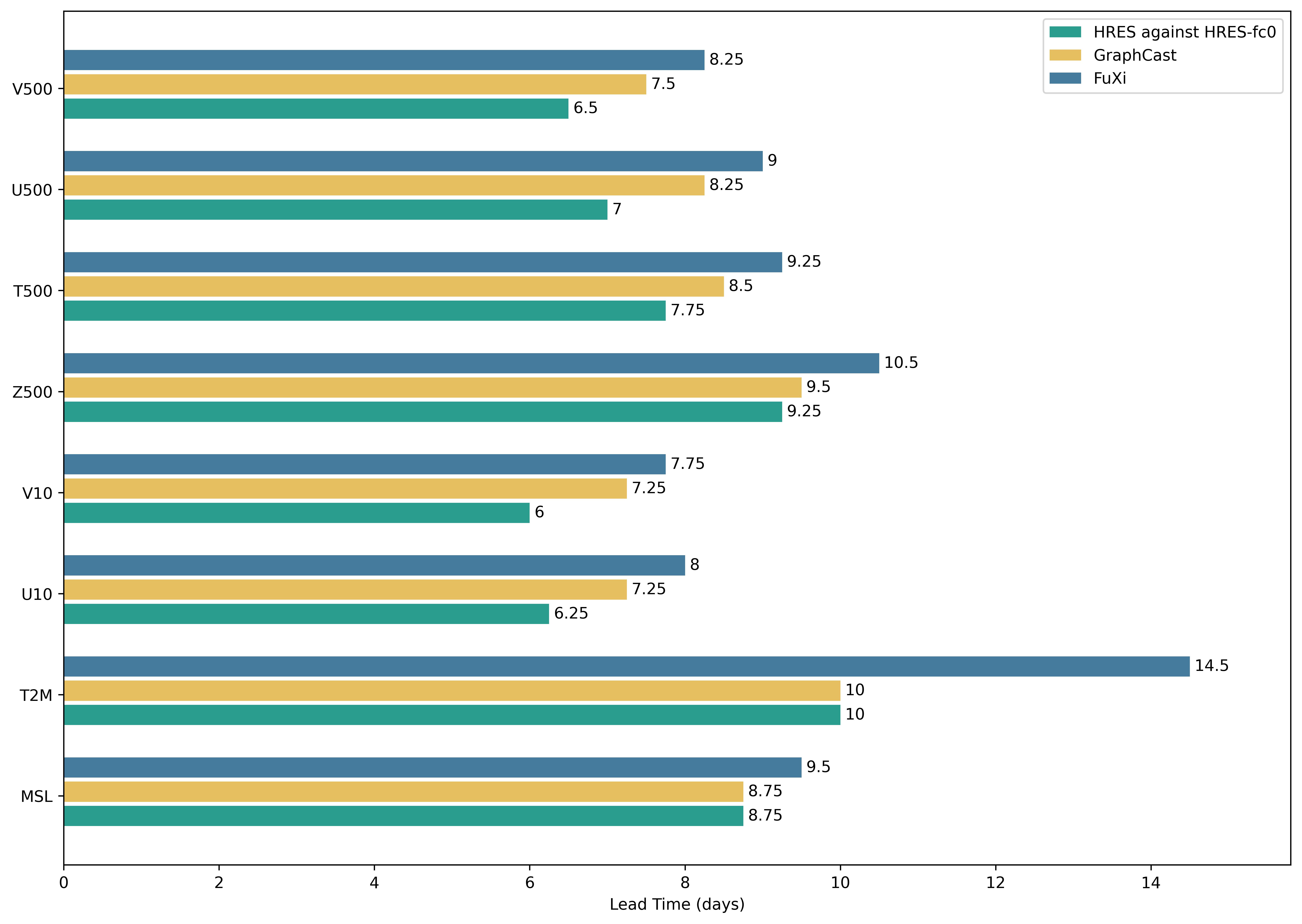}
    \caption{Skillful forecast lead times (ACC \gt 0.6) of ECMWF HRES, Graphcast, and FuXi for 4 surface variables (${MSL}$, ${T2M}$, ${U10}$, and ${V10}$) and 4 upper-air variables (${Z500}$, ${T500}$, ${U500}$, and ${V500}$) at 500 hPa pressure level.}
    \label{skillful_lead}    
\end{figure}

Figure \ref{skillful_lead} compares the skillful forecast lead time of ECMWF HRES, Graphcast, and FuXi for 4 surface variables: ${MSL}$, ${T2M}$, ${U10}$, and ${V10}$, as well as 4 upper-air variables at 500 hPa pressure level: (${Z500}$, ${T500}$, ${U500}$, and ${V500}$. FuXi improves the skillful lead time of all 8 variables show in the figure. The most significant improvement is for ${T2M}$, where FuXi increases the skilful lead time from 10 days by ECMWF HRES and Graphcast to 14.5 days for ${T2M}$. For ${Z500}$, the improvement is seen from 9.25 days by ECMWF HRES and 9.5 days by Graphcast to 10.5 days by FuXi.

\end{document}